# Gender Based Emotion Recognition System for Telugu Rural Dialects Using Hidden Markov Models

Prasad Reddy P.V.G.D [1]     Prasad A [2]     Srinivas Y [3]     Brahmaiah P [4]

**Abstract**—Automatic emotion recognition in speech is a research area with a wide range of applications in human interactions. The basic mathematical tool used for emotion recognition is Pattern recognition which involves three operations, namely, pre-processing, feature extraction and classification. This paper introduces a procedure for emotion recognition using Hidden Markov Models (HMM), which is used to divide five emotional states: anger, surprise, happiness, sadness and neutral state. The approach is based on standard speech recognition technology using hidden continuous markov model by selection of low level features and the design of the recognition system. Emotional Speech Database from Telugu Rural Dialects of Andhra Pradesh (TRDAP) was designed using several speaker's voices comprising the emotional states. The accuracy of recognizing five different emotions for both genders of classification is 80% for anger-emotion which is achieved by using the best combination of 39-dimensioanl feature vector for every frame (13 MFCCs, 13 Delta Coefficients and 13 Acceleration Coefficients) and a classifier using HMM. This outcome very much matches with that acquired with the same database with subjective evaluation by human judges. Both gender-dependent and gender-independent experiments are conducted on TRDAP emotional speech database.

**Index Terms**— Emotion Recognition, Support Vector Machines (SVM) classifier, Hidden Markov Models (HMM) classifier, Mel Frequency Cepstral Coefficients (MFCCs) and Telugu Rural Dialects of Andhra Pradesh (TRDAP).

—————————— ◆ ——————————

.
## 1.INTRODUCTION

Dealing with the speaker's emotion is one of the latest challenges in speech technologies. Three different factors can be easily recognized, namely, speech recognition in the presence of emotional speech, synthesis of emotional speech, and emotion recognition.

Speech is a time varying signal, which represents the underlying patterns of emotions. To capture these time varying underlying patterns, HMM can be effectively used. The way human beings react with the world and interact with it remains one of the greatest scientific challenges. The efficiency to recognize emotional state of a person is a most important factor for successful inter-personal social interaction. The emotional speech recognition system can be represented by the used features, the examined emotional categories, the methods to collect speech utterances, the languages, and the type of classifier used in the experiments. In this work HMM classifier and feature selection algorithm are used to classify five emotions from emotional speech and the results are presented using a confusion matrix for both gender-dependent and gender-independent survey.

Prasad A et al. [1] presented a work on Recognition of Telugu Rural Dialects by using Support Vector Machines (SVM). But the main disadvantage of SVM classifier is that it requires a fixed dimension input [2]. Also, SVM maximizes the margin which makes the emotions near the decision surface result in very uncertain classification, where there is at most 50% chance of classifier deciding either way. Hence in this paper HMM classifier is used to classify five emotions from emotional speech of TRDAP database. The overall experimental results disclose that the HMM classifier detects *anger* perfectly (80%), but confuses happiness with sadness, surprise and neutral.

## 2. EMOTION RECOGNITION SYSTEM

This comprises four modules: Emotional speech database, Feature extraction, HMM model and Recognized emotion output. The Emotional speech database used in this paper is from Telugu Rural Dialects of Andhra Pradesh for both genders. The speech is communicated in five emotional states: anger, surprise, happiness, sadness and neutral.

## 3. LOW LEVEL FEATURES OF EMOTION

The first problem that occurs when trying to build a HMM-based recognition framework is the discrimination of the features to be used. In this case, it is not enough that the feature carries information about the emotional state, but it must

———————————————

1. *Andhra University, Visakhapatnam, India.*
2. *Vignan University, Guntur, India.*
3. *VIIT, Visakhapatnam, India.*
4. *C.R. Atlantica Software Solutions, Hyderabad, India*

be relevant to the HMM structure as well. The main development of this limitation is that the features used must model the short-time emotion nullifying the use of global statistics estimated from the whole utterance. In this work, the raw features that could lead to statistical measures were selected, which are similar to those proposed in literature for emotion recognition [3].

Spectral measures were eliminated in our first approach to emotion recognition because they need complex frameworks to be characterized. This is so because the spectrum depends heavily on the phonetic content of the sentence. Pitch and energy also do, they also depend on extensive classes of sounds alone, rather than phonemes. Another reason for discarding spectral measures is the belief that their phonetic dependency would be a main deficiency for building language-independent emotion recognizers. Many named direct use of temporal while expelling silence related measures because the latter need a previous recognition step in order to get a phone / silence segmentation / recognition, increasing the complexity of the overall system.

The HMM structure along with a good choice of the pitch and energy features can provide a good representation of this kind of measures. Absolute values and long-term evolution of some parameters are avoided due to their dependency on factors that have nothing to do with the speaker's emotional state. For instance, the absolute value of energy reflects not only the intentional level, but also the gender and age of the speaker and the gain of the recording chain as well. On the other hand, whether a sentence is affirmative or interrogative and its length, will probably play a determinant role on the whole sentence contour of the pitch. For both energy and pitch, two kinds of temporal scopes have been considered; they are instantaneous values and syllabic contour. In the first case, raw low level analysis is performed on samples of 25 ms taken every 10 ms. In the second case, the same features are processed in order to capture their mean behavior in segments of 100 ms, that roughly correspond to the length of two phones.

### 3.1 ENERGY FEATURES

In order to model the instantaneous values of energy without relying on the absolute value of energy, the first and second derivatives of the logarithm of the mean energy are used in the frame. The acoustical meaning of these measures is related to the sharpness of the energy level, reflecting both the articulation speed and the dynamic range. Besides, the effects such as tremor, small frequent variations in voice intensity are also easily characterized by the instantaneous energy levels. Syllabic contour of energy is modeled by the first and second derivatives of the logarithm of the 8 kHz low pass filtered energy in the frame; in this case, the relative intensity of consecutive sounds is considered.

### 3.2 PITCH FEATURES

Pitch features present a similar behavior as energy ones. In this case, the global pitch which is heavily influenced by the speaker's nature and its global evolution along the utterance which depends on the sentence structure has been considered. The syllabic contour of pitch and its instantaneous levels to provide profitable information about the emotion are also considered.

In order to characterize instantaneous pitch, a simple auto-correlation analysis is performed at every frame. The maximum of the long term auto-correlation is determined and used to form five different parameters: the value of the maximum of the long term auto-correlation along with its first and second derivatives and the first and second derivatives of the logarithm of the pitch lag. The raw auto-correlation maximum is a measure of the harmony of voice. High values of this maximum imply a high periodicity in the speech waveform, while low values imply low or no periodicity. This feature allows us to discern between harsher styles such as anger or disgust and other more musical styles such as joy or surprise. Also, the first derivative of the logarithm of pitch lag represents the relative variation of pitch between frames.

In order to model the pitch lag, the position of the maximum of the long term auto-correlation, without further processing, is considered. Thus, it will present abundant errors, particularly pitch lag doubling and halving. These artifacts are characterized in the first derivative of the logarithm of the pitch lag as fixed constants. Thus, the first derivative of the logarithm of the pitch lag will help to detect those styles for which these effects are more frequent. They also help in the detection of jitter, the presence of fast fluctuations in the very short time values of pitch. The syllabic contour of pitch is characterized with a twice filtered version of the pitch lag estimated for instantaneous pitch: first the pitch lag is median filtered in order to remove artifacts from the estimation; secondly an 8 kHz low pass filter is applied to capture the syllabic contour. In this case, a much more precise estimation of pitch is used in order to capture the actual evolution of its values and not the frequency of errors in the estimation. The first and second derivatives of the smoothed pitch lag evolution, obtained this way, are then evaluated in order to represent the pitch evolution in segments of a few phones.

### 3.3 MEL-FREQUENCY CEPSTRAL COEFFICIENTS (MFCCS)

Any emotion identification system needs a robust acoustic-feature-extraction technique as a front-end block followed by an efficient modeling scheme for generalized representation of these features.

The MFCCs can represent the low frequency region more accurately than the high frequency region and hence then can capture formants which lie in the low frequency range and which characterize the vocal tract resonances. However, other formants can also lie above 1 kHz and these are not effectively captured by the larger spacing of filters in the higher frequency range [4]

All these facts suggest that any emotion recognition system, based on MFCCs, can possibly be improved. In this work, a new feature set from the speech signal which yields information that is complementary in nature to the human vocal tract is evaluated. This makes it very suitable to be used with a parallel classifier [5] to yield higher accuracy in emotion recognition problem.

MFCCs are coefficients that collectively make up an MFC, which are derived from a type of cepstral representation of the audio clip (a "spectrum-of-a-spectrum"). The difference between the cepstrum and the Mel-frequency cepstrum is that in the MFC, the frequency bands are equally spaced on the Mel scale, which approximates the human auditory system's response more closely than the linearly-spaced frequency bands used in the normal cepstrum.

Mel Frequency Cepstral Coefficients (MFCCs) are the dominant features used for speech recognition and to investigate their applicability to modeling music. In particular, two of the main assumptions of the process of forming MFCCs are examined for considering the Mel frequency scale to model the spectra, and the Discrete Cosine Transform (DCT) to decorrelate the Mel-spectral vectors.

Mel-Frequency Cepstral Coefficients (MFCCs) modeled on the human auditory system has been used as a standard acoustic feature set for emotion recognition applications. However, due to the structure of its filter bank, it captures vocal tract characteristics more effectively in the lower frequency regions.

The first assumption in the context of emotions discrimination, is examined and results show that the use of the Mel scale for modeling music is at least not harmful for this problem, the second assumption is examined on the basis vectors of the theoretically optimal transform to de-correlate emotion spectral vectors.

$X(f)$ per frame is used to calculate the Mel Frequency Cepstral Coefficients (MFCCs). The cepstral coefficients, which are the coefficients of the Fourier transform representation of the log magnitude spectrum, have been shown to be a more robust and reliable feature set for speech recognition than the Linear Predictive Coding (LPC) coefficients. Because of the sensitivity of the low order cepstral coefficients to overall spectral slope and the sensitivity of the high-order cepstral coefficients to noise, it had become a standard technique to weigh the cepstral coefficients by a tapered window so as to minimize these sensitivities.

In this work, a 39-dimensional feature vector (13 Mel Frequency Cepstral Coefficients, 13 Delta Coefficients, 13 Acceleration Coefficients) is generated per frame and this is used as the source feature. After the cepstral coefficients were generated, a Cepstral Mean Normalization (CMN) was done to get rid of the bias emotional signal present across the coefficients. This frequency warping can allow for better representation of sound.

MFCCs are commonly derived as follows [6]:
1. Take the Fourier transform of (a windowed excerpt of) a signal.
2. Map the powers of the spectrum obtained above onto the Mel scale, using triangular overlapping windows.
3. Take the logs of the powers at each of the Mel frequencies.
4. Take the discrete cosine transform of the list of Mel log powers, as if it were a signal.
5. The MFCCs are the amplitudes of the resulting spectrum [7].

MFCCs values are not very robust in the presence of additive noise, so it is necessary to raise log-Mel-amplitudes to a suitable power (around 2 or 3) before taking the DCT, which reduces the influence of low-energy components [3].

### 4. HIDDEN MARKOV MODELS (HMM)

The HMM is a doubly embedded stochastic process, which is not directly observable and has the capability of effectively modelling statistical variations in spectral features. In a variety of ways, HMMs can be used as probabilistic emotion recognition device [8] [9] [10]. HMM not only models the underlying speech sounds but also the temporal sequencing among the sounds. This temporal modelling is advantageous for emotion recognition.

In the training phase, one HMM for each emotion is obtained (i.e., parameters of model are estimated) using training feature vectors. The parameters of HMM [11] are:

- State-transition probability distribution, as represented by $A = [a_{ij}]$, where

$$a_{ij} = P(q_{t+1} = j / q_t = i) \qquad 1 \leq i, j \leq N \quad (4.1)$$

This defines the probability of transition from state i to j at time t.

- Observation symbol probability distribution, as given by $B = [b_j(k)]$ in which

$$b_j(k) = P(o_t = v_k / q_t = j) \qquad 1 \leq k \leq M \quad (4.2)$$

This defines the symbol distribution in state j, j = 1,2,…N

- The initial state distribution, as given by $\Pi = [\pi]$, where

$$\pi_i = p(q_1 = i) \qquad 1 \leq i \leq N \quad (4.3)$$

where, N is the total number of states, and $q_t$ is the state at time t, M is the number of distinct observation symbols per state, and 0t is the observation symbol at time $t$. In testing phase, $P(O/\lambda)$ for each model is calculated, where $O = (o_1 o_2 o_3 ...... o_n)$. Here the goal is to find out the probability for a given model to which the test utterance belongs. The emotion is recognized by the highest likelihood value. The other characteristics are summarized as follows:

- The output probability of the HMM model is in terms of multiplication of several probabilities. Therefore the scores are very small. Efficient scaling methods have to be derived to overcome this computational overhead.
- HMM is also a statistical approach, and hence requires large amount of training data for effective estimate of the model parameters.
- The system performance degrades when training and testing environments differ.

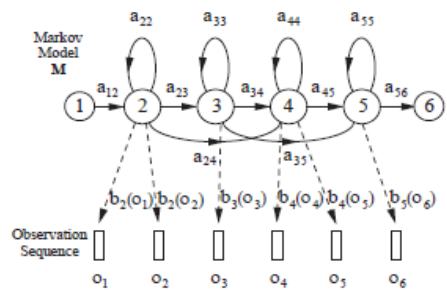

Figure 3.1: The Markov Generation Model

## 5. EXPERIMENTAL EVALUATION

The emotions, from a given set of emotion data are captured and LPC residuals are applied. Feature extraction is done using the best combination of 39 dimensional feature vectors for every frame (13 MFCC's, 13 Delta Coefficients and 13 Acceleration Coefficients) to capture the emotions information. The TRDAP corpus of emotion data has been used to evaluate the emotion recognition for its emotive content. 200 speeches with maximum of 30 seconds of speech data for training and minimum of 1 second of data for testing are considered.

## 6. RESULTS AND DISCUSSION

The results of classification obtained from TRDAP database of both gender-dependent and gender-independent features are combined to produce more accurate results. The regions commonly identified in both the classifications are now highlighted. In this work, the HMM classifier is considered along with a novel parameter vector called 39-dimensional vector, to distinguish the five emotional states. From these values, emotion features like anger, surprise, happiness, sadness and neutral state dissimilarities are obtained. The emotional features that are extracted are shown in the following tables.

### 6.1 CLASSIFIED VALUES OBTAINED FROM GENDER INDEPENDENT SURVEYS:

| Stimulation | Recognized Emotions (%) | | | | |
|---|---|---|---|---|---|
| | anger | surprise | happiness | sadness | neutral |
| anger | **80** | 0 | 10 | 0 | 10 |
| surprise | 0 | **70** | 0 | 20 | 10 |
| happiness | 10 | 10 | **80** | 0 | 0 |
| sadness | 0 | 20 | 10 | **70** | 0 |
| neutral | 10 | 0 | 0 | 10 | **80** |

Table 1: Confusion matrix indicating different

Gender-independent emotions using HMM
It can be noted from the table that in gender-independent cases, the HMM classifier recognizes anger, happiness and neutral more accurately with 80 % compared to other emotions. The next is sadness with 70 %. The mis-classification has values with a maximum of 20 % and hence are marked by the figures of correct classification.

### 6.2 CLASSIFIED VALUES OBTAINED FROM GENDER DEPENDENT SURVEYS:

| Stimulation | Recognized Emotions (%) | | | | |
|---|---|---|---|---|---|
| | anger | surprise | happiness | sadness | neutral |
| anger | **80** | 0 | 10 | 10 | 0 |
| surprise | 10 | **70** | 0 | 20 | 0 |
| happiness | 10 | 10 | **70** | 0 | 10 |
| sadness | 10 | 0 | 10 | **60** | 20 |
| neutral | 0 | 10 | 0 | 20 | **70** |

Table 2: Confusion matrix indicating different Gender dependent emotions of Male using HMM

In the experiments conducted with gender-dependent (Male) cases, the HMM classifier did not do as perfectly as with gender-independent cases. While it recognized anger with 80 % accuracy, the other four emotions have 60 % - 70 %, the one emotion with the lowest 60 % is sadness and the other three have 70 %. Again, the mis-classification figures are small and do not affect the perfect classification.

| Stimulation | Recognized Emotions (%) | | | | |
|---|---|---|---|---|---|
| | anger | surprise | happiness | sadness | neutral |
| anger | **80** | 0 | 10 | 0 | 10 |
| surprise | 0 | **70** | 10 | 20 | 0 |
| happiness | 10 | 10 | **70** | 0 | 10 |
| sadness | 10 | 10 | 0 | **60** | 20 |
| neutral | 10 | 0 | 10 | 0 | **80** |

Table 3: Confusion matrix indicating different Gender dependent emotions of Female using HMM

As can be seen from Table 3, for gender-dependent (Female) cases, the results are similar to those with male cases, although here it is slightly better.

### 6. CONCLUSION

In this paper, Hidden Markov Models (HMM) classifier, with 39 Coefficients (13 MFCCs, 13 Delta Coefficients and 13 Acceleration Coefficients), are considered as features for emotion recognition system. For this, the emotions from Telugu Rural Dialects of Andhra Pradesh (TRDAP) database of both genders, male and female, are used to detect five basic emotions. The results of recognized emotions are presented in a confusion matrix based on samples collected from both the genders. As indicated in the experimental results, a high degree of accuracy of 80% in recognizing anger emotion and more than 60% accuracy in other emotions are achieved. The HMM classifier achieved a better performance in recognizing anger emotion in all the cases. The classifier did significantly well with gender-independent cases.